\documentclass[twocolumn,showpacs,preprintnumbers,amsmath,amssymb,10pt, aps, prl]{revtex4-1}

\usepackage{graphicx}
\usepackage{amsmath}

\newcommand{\Fig}[2][]{Fig.~\ref{#2}#1}

\begin{document}

\title{Quantum frequency conversion between infrared and ultraviolet}

\author{Helge R\"utz}
\email[]{helge.ruetz@uni-paderborn.de}
\author{Kai-Hong Luo}
\author{Hubertus Suche}
\author{Christine Silberhorn}

\affiliation{Integrated Quantum Optics, Applied Physics, University of Paderborn, Warburger Str.~100, 33098 Paderborn, Germany}
\date{\today}

\begin{abstract}

We report on the implementation of quantum frequency conversion (QFC) between infrared (IR) and ultraviolet (UV) wavelengths by using single-stage upconversion in a periodically poled KTP waveguide. Due to the monolithic waveguide design, we manage to transfer a telecommunication band input photon to the wavelength of the ionic dipole transition of Yb${}^{+}$ at 369.5 nm. The external (internal) conversion efficiency is around 5\% (10\%). The high energy pump used in this converter introduces a spontaneous parametric downconversion (SPDC) process, which is a cause for noise in the UV mode. Using this SPDC process, we show that the converter preserves non-classical correlations in the upconversion process, rendering this miniaturized interface a source for quantum states of light in the UV.

\end{abstract}

%
%
%
%
\maketitle

Quantum networks and long distance quantum communication rely on the faithful transfer and manipulation of quantum states. 
Because a single quantum system does not necessarily incorporate all the benefits needed, a hybrid system~\cite{Kimble2008, Schleier-Smith2016} with different nodes operating at dissimilar frequencies may be used to perform each task at its optimal frequency.
The process of quantum frequency conversion (QFC)~\cite{Kumar1992, Tanzilli2005} has been established as a means to bridge the gap between differing frequencies while keeping the quantum correlations intact.
On the one side of that gap, telecommunications bands in the infrared spectral region have consensually been identified as the optimal wavelengths for quantum state transfer because of low loss in optical fibers and a multitude of experimental studies have shown QFC 
from~\cite{Tanzilli2005,  Honjo2007,  Rakher2010,  Rakher2011,  Vollmer2014,  Liu2015,  Zhou2016,  Baune2016} and 
to~\cite{Ikuta2011,  Zaske2012,  DeGreve2012,  Pelc2012,  Albrecht2014,  Yu2015} the telecommunications bands, c.f.~\Fig[~(a)]{fig:conceptimage}.
Looking at the other side of the gap, one finds that QFC experiments have so far mainly focused on convenient laser wavelengths or transitions in the red/near-infrared spectral region.
However, high fidelity photonic state manipulation strongly benefits from high energy transitions and indeed, most of the beneficial ionic transitions are situated at ultraviolet (UV) wavelengths.
Most prominently, the Ytterbium- (Yb$^+$) transition at $369.5\,\mathrm{nm}$ constitutes an almost ideal 2-level quantum system due to the Yb$^+$-ion's specific electronic structure~\cite{Olmschenk2010}. 
It has been shown that the $S_{1/2} \rightarrow P_{1/2}$ transition can act as a photonic interface for efficient and long lived storage of quantum bits~\cite{Olmschenk2007}, and the Yb-ion proves to be useful for quantum computing~\cite{Hannemann2002, Olmschenk2010} and fundamental studies of light-matter interactions~\cite{Maiwald2012}. 
While this ion is thus an ideal system for the manipulation of quantum states, its application for quantum networks and long distance quantum communication in a hybrid system is conditioned on the possibility to connect it to the optimal fiber transmission window (c.f. fiber attenuation in~\Fig[~(c)]{fig:conceptimage}).
This is a challenging venture and involves significant engineering efforts because of the large energy gap between in- and output, as well as the properties of nonlinear materials at UV wavelengths. 
In contrast to a proposal to use a two-stage cavity system~\cite{Clark2011}, we have recently 
demonstrated a classical upconversion process 
in a rubidium doped periodically poled potassium titanyl phosphate (PPKTP) waveguide~\cite{Ruetz2016}, which provides the basis for QFC. 

\begin{figure}
\begin{center}
\includegraphics*[width=0.99\columnwidth]{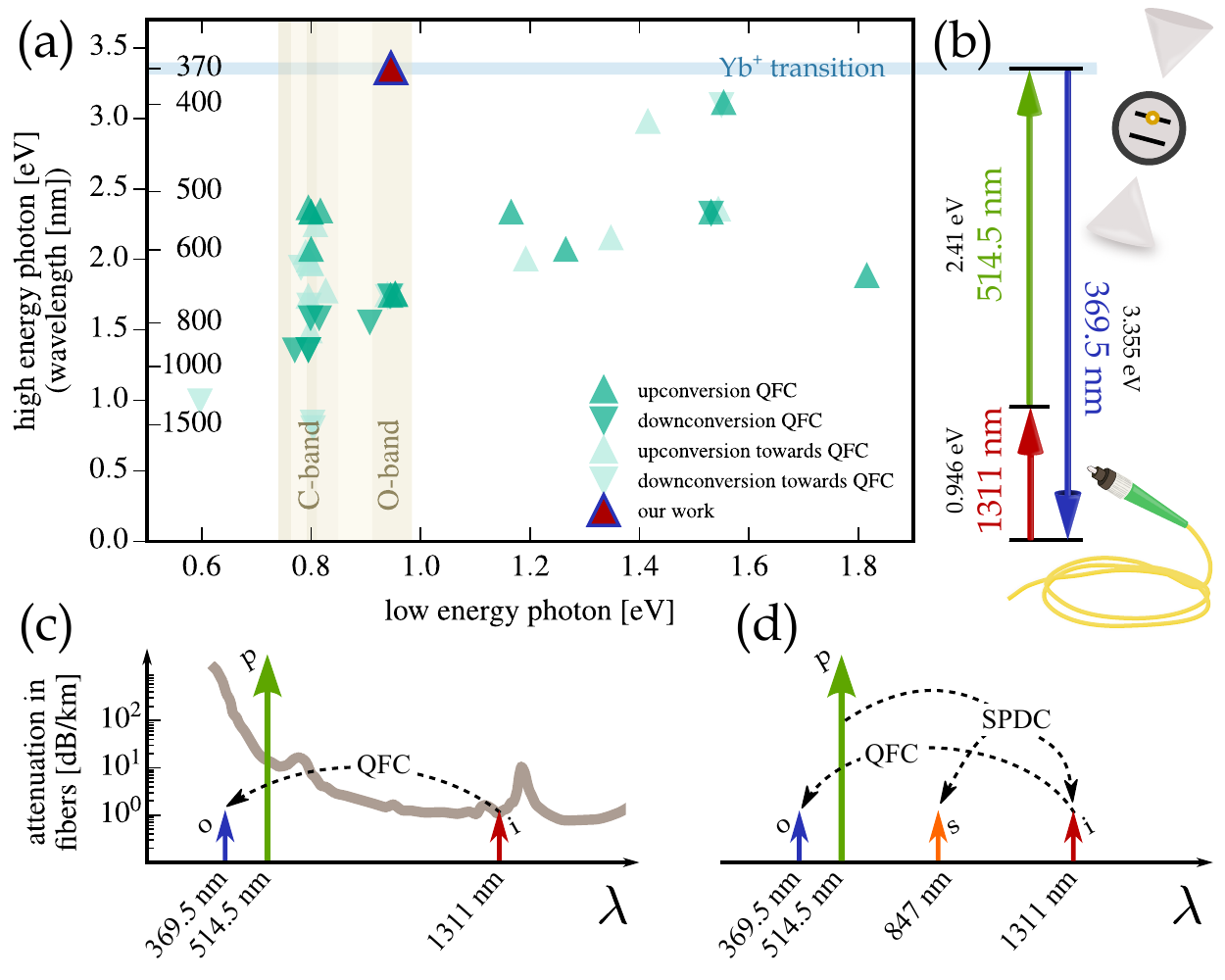} 
\end{center}
\caption{\label{fig:conceptimage}
Quantum frequency conversion between telecommunication band and UV. The conversion to the UV wavelength of $369.5\,\mathrm{nm}$ (dark edged triangle) is contrasted to state-of-the-art quantum frequency conversion (dark triangle)~\cite{Kumar1992,  Rakher2010,  Ates2012,  Zaske2012,  Ikuta2011,  Tanzilli2005,  Vollmer2014,  Kong2014,  Honjo2007,  Ramelow2012,  Rakher2011,  McGuinness2010,  DeGreve2012,  Yu2015,  Pelc2012,  Donohue2014,  Albrecht2014,  Radnaev2010,  Zhou2016,  Liu2015} as well as important work towards QFC (light triangle)~\cite{Vandevender2004,  Albota2004,  Thew2006,  Langrock2005,  Langrock2005,  Kuo2013,  Hill2011,  TurnerFoster2010,  Roussev2004,  Clark2013,  Takesue2010,  Pelc2011,  Curtz2010,  Brecht2014,  Steinlechner2016,  Gu2012,  Giorgi2003,  Ding2010,  Zaske2011,  Ikuta2014,  Cheng2015} in a plot that has the input and output photon energy on its axes (a).
The concept presented on an energy (b) and wavelength (c) scale. For comparison, the attenuation in optical fibers~\cite{Langner2007}, which makes direct transmission of UV light impossible, is shown (c). The concept of a cascaded SPDC/QFC process used for QFC to the UV is depicted in (d). }
\end{figure}%
%

In this letter, we report on the QFC between infrared and UV for single photon states. More specifically we show QFC between the telecommunications O-band (around $1310\,\mathrm{nm}$) and the wavelength of the Yb${}^{+}$ transition at $369.5\,\mathrm{nm}$, bridging an energy gap larger than $2.4\,\mathrm{eV}$ ($580\,\mathrm{THz}$) and directly matching the $S_{1/2} \rightarrow P_{1/2}$ dipole transition in Yb$^+$. 
By converting single photon level light we show an external (internal) efficiency above 5\% (10\%).
Using an intrinsic spontaneous parametric downconversion (SPDC) process, we verify the preservation of non-classicality between the in- and output mode. This SPDC process on the one hand acts as a source of noise in the conversion process, but also allows -- in conjunction with the upconversion process -- to produce quantum states of light in the UV, which would otherwise require a pump at even shorter wavelengths.

%

The concept of this QFC on an energy and a wavelength scale is depicted in~\Fig[~(b)~and~(c)]{fig:conceptimage}, respectively. The QFC-Hamiltonian is given by~\cite{Kumar1990}
\begin{equation}\label{eqn:qfc}
\hat{\mathcal{H}}_\mathrm{QFC} = \mathrm{i}\hbar\kappa A_\mathrm{p} \hat{a}_\mathrm{i} \hat{a}^{\dagger}_\mathrm{o} + \mathrm{h.c.}
\end{equation}
where $\hat{a}_\mathrm{\{i,o\}}$ are annihilation operators for the input (i) and output (o) mode, $A_\mathrm{p}$ is the classically treated pump field, and $\kappa$ is the coupling constant which incorporates the nonlinearity of the material as well as the transverse overlap of the light fields. 
In principle the QFC process is noiseless and can work with unit efficiency, a fact that can be anticipated from the beamsplitter-like form of eqn.(\ref{eqn:qfc}).
In practice however, losses and detrimental effects caused by the strong pump field limit the attainable efficiency. 
\begin{figure}
\begin{center}
\includegraphics*[height=1.809in]{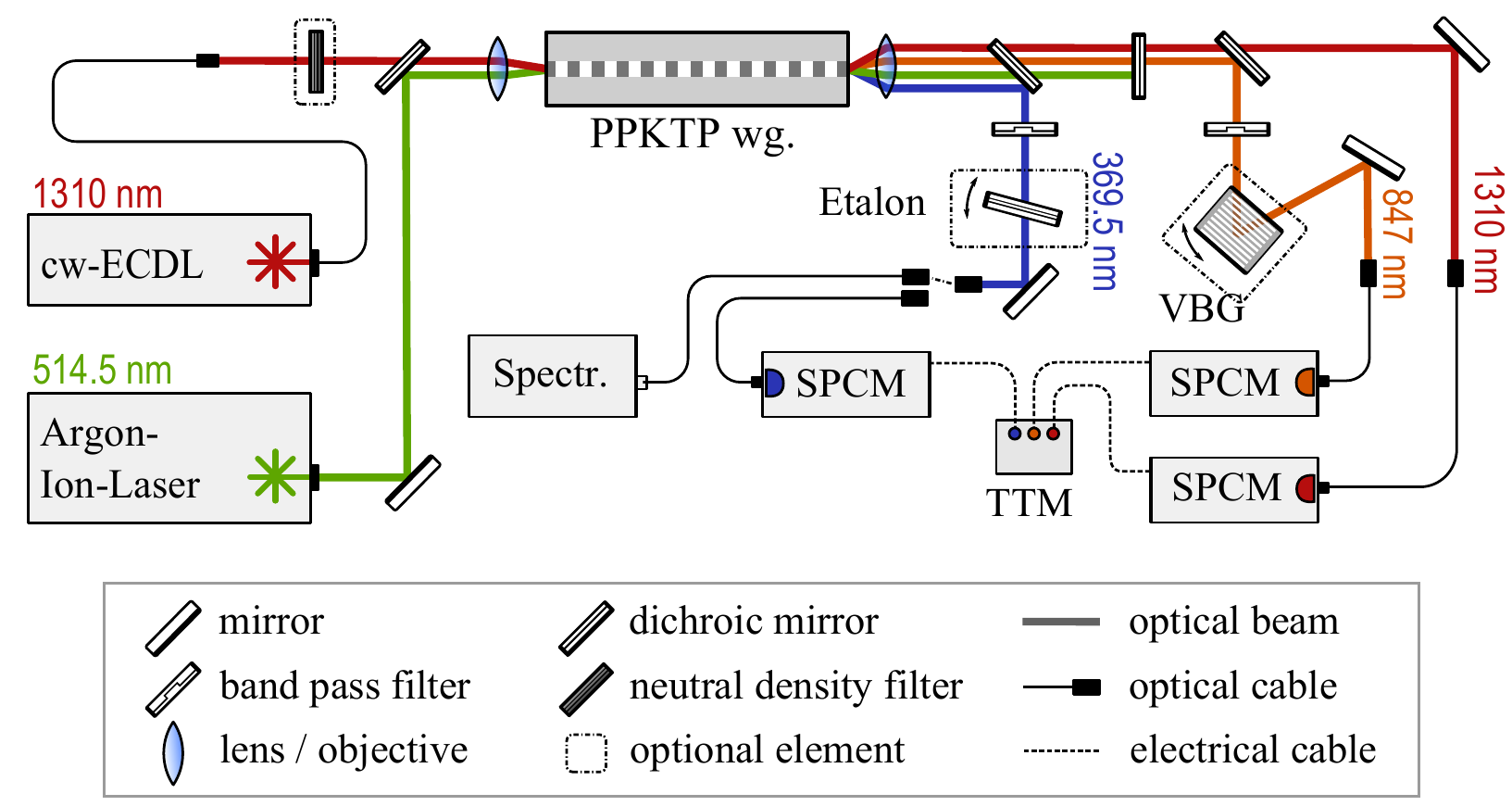}
\end{center}
\caption{\label{KTPSetup}
Schematic of the experimental setup. Details are described in the text. The following abbreviations are used; VBG: volume bragg grating,  
SPCM: Single photon counting module, Spectr.: sensitive spectrometer, TTM: time tagging module. Elements surrounded by doted lines are optionally placed into the beam path.}
\end{figure}%
%
The noise in the converter we report on here is caused by a SPDC process as sketched in~\Fig[~(d)]{fig:conceptimage}. Here, the strong pump decays into a pair of signal (s) and idler (i) photons, where the idler mode is identical to the input mode of the QFC process according to
\begin{equation}\label{eqn:spdc}
\hat{\mathcal{H}}_\mathrm{SPDC} = \mathrm{i}\hbar\gamma A_\mathrm{p} \hat{a}^{\dagger}_\mathrm{s} \hat{a}^{\dagger}_\mathrm{i} + \mathrm{h.c.} ,
 \end{equation}
where $\gamma$ is the coupling constant for the SPDC process.
The produced SPDC state exhibits strong nonclassical correlations~\cite{Gerry2004} and these correlations are preserved in the frequency conversion process as we verify experimentally in this paper.
Theoretically, the output state of the generation and successive conversion over time $t$ can be calculated by the evolution 
$\hat{U} =  \hat{U}_\mathrm{QFC}\hat{U}_\mathrm{SPDC}$ 
with 
$\hat{U}_\mathrm{QFC} = \exp{( \mathrm{i}t\hat{\mathcal{H}}_\mathrm{QFC} / \hbar )}$ 
and 
$\hat{U}_\mathrm{SPDC} = \exp{( \mathrm{i}t\hat{\mathcal{H}}_\mathrm{SPDC} / \hbar )}$, 
\begin{equation}\label{eqn:evo}
\hat{U} \left|0_\mathrm{s}, 0_\mathrm{i}, 0_\mathrm{o}\right\rangle 
\sim \gamma A_\mathrm{p}t \left|1_\mathrm{s}, 1_\mathrm{i}, 0_\mathrm{o}\right\rangle + \gamma\kappa A_\mathrm{p}^2 t^2  \left|1_\mathrm{s}, 0_\mathrm{i}, 1_\mathrm{o}\right\rangle ,
\end{equation}
showing quantum correlations between a signal and an output mode in quadratic dependence on the pump field.  
%
\begin{figure}
\begin{center}
\includegraphics*[height=1.809in]{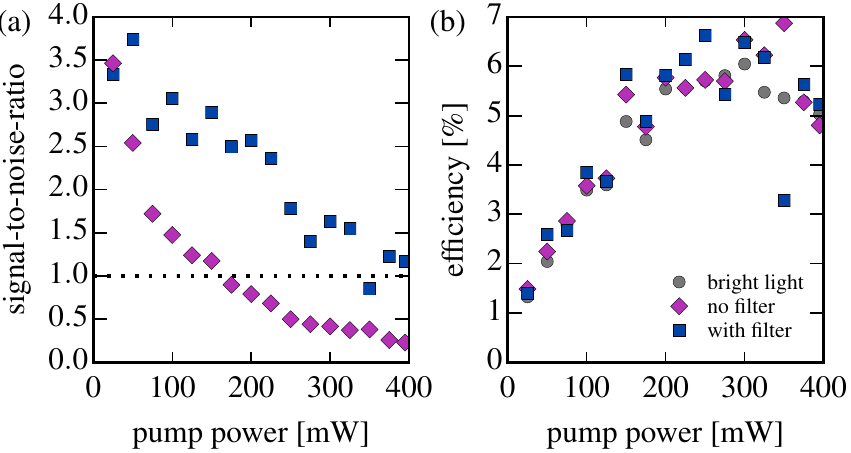}
\end{center}
\caption{\label{fig:KTPConversion}
Upconversion at the single photon level. Rates are measured as the number of counts within one second, averaged over 10 successive measurements. (a) Signal-to-noise ratio with and without etalon filtering (see text). (b) Conversion efficiency of bright ($22\,\mu\mathrm{W}$) and weak ($1\,\mathrm{pW}$) telecommunications band input light.}
\end{figure}%

The implementation of QFC between the IR and the UV is based on a type-0 sum frequency generation (SFG) process with a strong pump at $514.5\,\mathrm{nm}$ in a PPKTP waveguide (AdvR Inc.). The waveguide has a length of $L = 9.6\,\mathrm{mm}$ and the nominal poling period is $\Lambda = 2.535\,\mu\mathrm{m}$.

The experimental setup to study this QFC is shown in~\Fig{KTPSetup}. 
The waveguide chip is pumped with a continuous-wave argon ion laser 
in single longitudinal and spatial mode operation, while 
the IR input beam is produced by a tunable external cavity diode laser.
Both beams are overlapped on a dichroic mirror and then launched into the waveguide, where care must be taken to excite the fundamental mode. 
The sample is stabilized with an accuracy of $\pm 4\,\mathrm{mK}$ around room temperature to obtain quasi-phase-matching (QPM) such that the SFG ($1311\,\mathrm{nm} + 514.5\,\mathrm{nm} \rightarrow 369.5\,\mathrm{nm}$) takes place inside the waveguide. The exact QPM wavelength is tunable by temperature, as well as pump power, as discussed in our classical characterization~\cite{Ruetz2016}.
Behind the waveguide, we use a dichroic mirror to separate the UV light from the pump and input light, then several successive dielectric filters 
(with a bandwidth of $\sim 6\,\mathrm{nm}$ and a cumulative optical density above 22 at the pump wavelength) centered at $370\,\mathrm{nm}$ to filter out the remaining pump light.

First, we investigate the conversion efficiency on the single photon level and apply narrow filtering to reduce noise contributions. 
To this end, 
the generated UV light is coupled to a a blue enhanced single photon counting module (SPCM)
with a dark count rate of $13\,\mathrm{Hz}$. 
As input we use CW light 
with a photon flux of $I = 6\,\mathrm{MHz}$ ($\sim 1\,\mathrm{pW}$), which is compatible to state-of-the-art single photon sources~\cite{Strauf2007, Pomarico2012}.
For evaluating the signal-to-noise ratio (SNR), $(S+N)/N$, we record the output count rate $S$ as the number of counts within one second, averaged over 10 successive measurements, as well as the noise rate $N$, i.e. the rate without any input. 
In order to increase the SNR, narrow filtering is applied, i.e. we insert a homemade etalon with a free spectral range of $340\,\mathrm{GHz}$ ($\sim$ $150\,\mathrm{pm}$ at $370\,\mathrm{nm}$ wavelength) and a nominal bandwidth of $5.5\,\mathrm{GHz}$ ($2.5\,\mathrm{pm}$) into the beam path. 
In~\Fig[~(a)]{fig:KTPConversion} we plot the measured SNR as a function of the pump power for the etalon filtered (non-filtered) case as squares (diamonds) and find an SNR above 2 at pump powers up to $200\,\mathrm{mW}$ when using narrow filtering.

We define the external conversion efficiency $\eta_\mathrm{ext}$ as the number of converted photons exiting the nonlinear crystal $\left\langle n\right\rangle_\mathrm{out}$ divided by the number of input photons in front of it $\left\langle n\right\rangle_\mathrm{in}$, 
$\eta_\mathrm{ext} = \left\langle n\right\rangle_\mathrm{out} / \left\langle n\right\rangle_\mathrm{in} = (S-N)/(I\cdot \eta_\mathrm{loss})$, where $\eta_\mathrm{loss}$ accounts for optical losses outside the waveguide chip ($\sim\!22\%$), fiber coupling ($\sim\!69\%$), the detection efficiency ($\sim\!14\%$) and, where appropriate, for the etalon transmission ($\sim\!50\%$). 
\Fig[~(b)]{fig:KTPConversion} shows
the pump power dependent conversion efficiencies at the single photon level in the filtered (squares) and unfiltered (diamonds) case.
The external conversion efficiency is $\eta_\mathrm{ext} = 5.5\,\%$ at $200\,\mathrm{mW}$ pump power; in this case an SNR above 2 is achieved at the detector. Considering the mode matching into the waveguide based on transmission measurements~\cite{Ruetz2016}, we estimate an internal conversion efficiency of $\eta_\mathrm{int} = 10.5\,\%$. For both, the filtered and unfiltered case, the same conversion efficiency is obtained within the range of experimental repeatability. This is in good agreement to measurements using bright input light around $20\,\mu\mathrm{W}$~\cite{Ruetz2016} as an input (circles in~\Fig[~(b)]{fig:KTPConversion}) showing faithful upconversion over 7 orders of magnitude of input signal.
While higher efficiencies can be observed at higher pump powers, using the converter at a pump power of $200\,\mathrm{mW}$ seems to be a good compromise between optimal conversion efficiency and acceptable SNR. 
This is especially appropriate since the conversion efficiency saturates at elevated pump powers, which we attribute to a pump power dependent UV absorption. Using a pulsed pump may significantly reduce this saturation effect, such that an internal efficiency up to $30\,\%$ could potentially be reached for the same pump peak power levels~\cite{Ruetz2016}.
Note that due to the symmetry of the QFC-Hamiltonian, the inverse process of downconverting a UV photon to the telecommunication band in the same device can be expected to work with a similar internal efficiency -- with the external efficiency solemnly limited by the technicality of diffraction limited modematching to the UV waveguide mode.
\begin{figure}
\begin{center}
\includegraphics*[viewport=8 9 235 121]{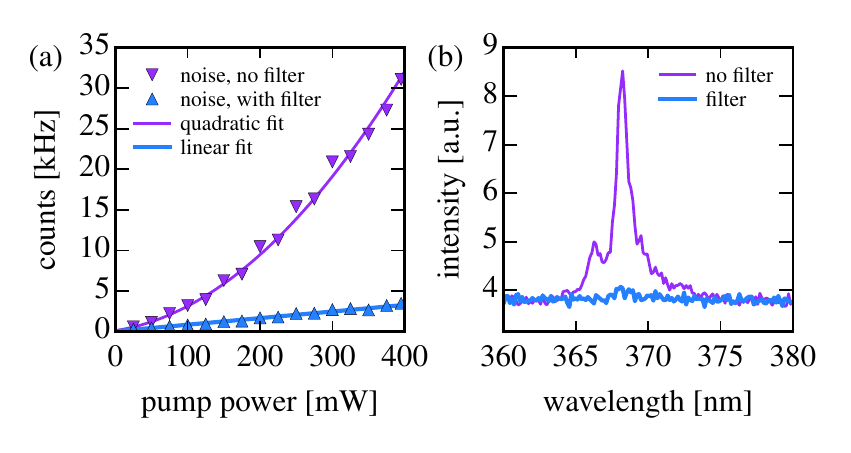} 
\end{center}
\caption{\label{fig:background} Noise in single photon level upconversion. (a) Count rates at the wavelength of the upconverted light around $370\,\mathrm{nm}$ without input. (b) Spectrally resolved output noise with and without narrow etalon filtering. }
\end{figure}%
%

We now consider the noise background stemming from SPDC in more detail.
While in principle, noise in QFC can also be caused by spontaneous Raman scattering (SRS) of the strong pump~\cite{Pelc2010} it is not to be expected in our process as the pump is spectrally well separated from the output ($>0.9\,\mathrm{eV}$).
In our QFC the noise is mainly due to a cascaded SPDC/SFG process, as sketched in~\Fig[~(d)]{fig:conceptimage}. In the first stage of this cascaded process, pump photons non-degenerately decay into two daughter photons, one of which, \textit{idler} (i), is at the wavelength band of the input light around $1311\,\mathrm{nm}$, the other, \textit{signal} (s), keeps energy conservation. 
Subsequently, due to the QPM of the our SFG process, the photon produced at the input wavelength is converted to the UV output (o) and appears as noise inside the detection band. As each of those processes is operated at low efficiency, they show a linear behavior as a function of the pump power. The cascaded process therefore evolves quadratically with pump power, see eqn.~(\ref{eqn:evo}). Such a process can always happen if the pump for the QFC process is located at a wavelength shorter than the input wavelength~\cite{Pelc2010, Maring2014}.
 
The count rates without input in the filtered (unfiltered) case are shown in~\Fig[~(a)]{fig:background} as upward (downward) triangles.
In the unfiltered case the count rate shows a quadratic dependence on the pump power, which is in agreement with other reported converters in the visible range~\cite{Maring2014} and can be anticipated from eqn.~(\ref{eqn:evo}).  
The count rate with narrow filtering is significantly reduced and appears linear as a function of pump power. 
The output
from the waveguide is then coupled to a spectrometer system 
with a resolution of about $0.15\,\mathrm{nm}$. \Fig[~(b)]{fig:background} shows the spectrum when only pump light is coupled to the waveguide. When no etalon is used, it shows a pronounced broadband peak and its shape resembles the phasematching curve of the SFG process~\cite{Ruetz2016}. 
Using the etalon almost completely eliminates the peak. The remaining noise is thus very broadband and could be easily filtered out using etalon cascades. It should also be mentioned that in the envisioned usage case the atomic transition of the ion   
can itself act as a narrow filter to increase the SNR, i.e. within its $20\,\mathrm{MHz}$ bandwidth a noise count rate of only $1.3\,\mathrm{Hz}$ would be expected. 

%
%
%
Finally, we measure coincidence events between the wavelengths in question, thereby showing the parametric nature of the noise and the preservation of quantum features in the frequency conversion process.
\begin{figure}
  \begin{center}
    \includegraphics[viewport= 9 10 236 121]{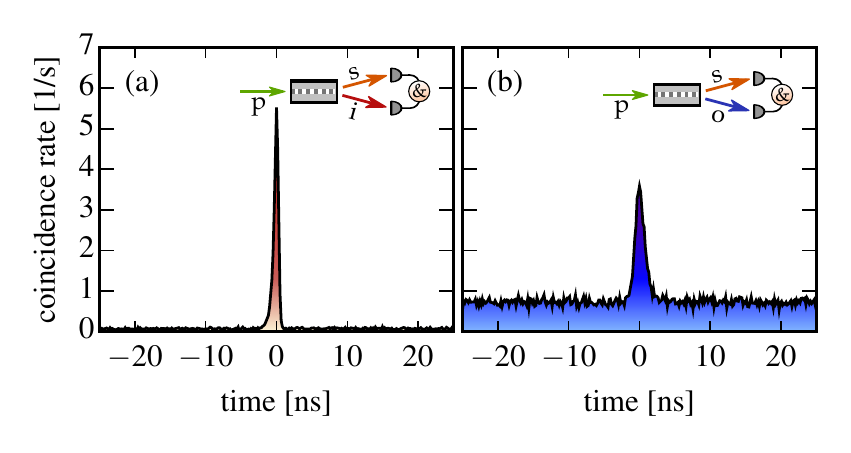} 
	\end{center}
    \caption{\label{fig:coinc} Coincidence measurements.
		(a) Coincidence measurement between signal around $847\,\mathrm{nm}$ and infrared. A pump power of $400\,\mu\mathrm{W}$, time binning of $165\,\mathrm{ps}$ and total measurement time of $230\,\mathrm{s}$ are used.
		(b) Coincidence measurement between signal around $847\,\mathrm{nm}$ and UV output. A pump power of $200\,\mathrm{mW}$ and total measurement time of $136\,\mathrm{s}$ are used.
		}
  \end{figure} %
By evaluating the cross-correlation between the mode at signal and idler, as well as between signal and the converted UV mode we show the non-classicality of the state and its preservation during frequency conversion. 
We pump the waveguide at $\sim 400\,\mu\mathrm{W}$ pump power and separate the near infrared from the telecommunications O-band light. The latter is coupled to a free-running SPCM. 
The light around $850\,\mathrm{nm}$ is filtered using a bandpass filter (FWHM $\sim 4\,\mathrm{nm}$) and launched to another SPCM. 
Detector events are recorded by means of a time tagging module. 
A histogram of the time difference $\tau$ between events in both channels is plotted in \Fig[~(a)]{fig:coinc}. A singular peak in the coincidence rate is observed, stemming from correlated SPDC photons in the two modes. 

The cross-correlation function $g^{(2)}_\mathrm{s,i}$ is given by $g^{(2)}_\mathrm{s,i} = p_\mathrm{s,i} / (p_\mathrm{s}p_\mathrm{i})$, where $p_\mathrm{s,i}$ is the probability for a coincidence event and $p_\mathrm{s}$ and $p_\mathrm{i}$ are the probabilities of detecting a photon in either mode~\cite{Clauser1974}. In practice, this ratio is calculated by 
$ g^{(2)}_\mathrm{s,i} = r_\mathrm{s,i}(\tau =0) / r_\mathrm{s,i}(\tau \neq 0) $,
dividing the coincidence rate at the peak's position $r_\mathrm{s,i}(\tau =0)$ by the coincidence rate outside the coincidence window $r_\mathrm{s,i}(\tau \neq 0)$.
The Cauchy-Schwarz inequality $g^{(2)}_\mathrm{s,i} \leq \sqrt{ g^{(2)}_\mathrm{s}(0)\cdot g^{(2)}_\mathrm{i}(0) }$ ~\cite{Reid1986}, with $g^{(2)}_\mathrm{s}(0)$ and $g^{(2)}_\mathrm{i}(0)$ being the auto-correlation function of the two fields at zero time delay, serves as a criterion for classicality. 
The perfect thermal state would show an auto-correlation of 2, imperfections and noise only lowering this value towards 1, limiting $g^{(2)}_\mathrm{s,i} \leq 2$ for classical states.
Here, we obtain $g^{(2)}_\mathrm{s,i} = 78\pm 20$, indicating the expected non-classicality of the SPDC state.

To measure the cross-correlation function $g^{(2)}_\mathrm{s,o}$ between the signal and the output state we increase the pump power to $200\,\mathrm{mW}$, high enough for efficient conversion. 
In the UV beam path we use the $6\,\mathrm{nm}$ filters, but no etalon and couple the light to the SPCM. In the red beam path we install a volume bragg grating (FWHM $\sim 1\,\mathrm{nm}$). This is necessary because the high pump power reduces the value of the cross-correlation function as more luminescence noise couples to the modes. Still, a clear peak is observed in the coincidence rates between signal and output modes, which is shown in~\Fig[~(b)]{fig:coinc}. 
The cross-correlation is $g^{(2)}_\mathrm{s,o} = 4.9\pm 0.5 \nleq 2$, violating the Cauchy-Schwarz classicality criterion by more than 5 standard deviations. 
Note that we would expect an even higher violation using external SPDC photons from a narrow-band source~\cite{Luo2015} due to the possibility of improved filtering of the near-infrared photons.
The preservation of correlations together with the SNR above 2 make this device directly deployable for time-bin qubit conversion~\cite{Tanzilli2005, Ikuta2011}, while polarization qubits would require an appropriate multiplexing scheme (e.g. Sagnac-loop).

In conclusion, we have implemented QFC between IR and UV wavelengths based on SFG in a PPKTP waveguide.
Using a fixed single-mode pump at $514.5\mathrm{nm}$, the device allows to interface the telecommunication band at $1311\,\mathrm{nm}$ to the Yb${}^{+}$ transition at $369.5\,\mathrm{nm}$ with an external (internal) efficiency of $\eta_\mathrm{ext} = 5.5\ \%$ ($\eta_\mathrm{int} = 10.5\ \%$). 
The device retains its conversion properties on the single-photon-level and is quantum state preserving, which is shown by converting an intrinsic SPDC state.
This intrinsic SPDC process is the main noise contribution for the conversion. However, strong filtering can notably limit its influence and an even higher noise suppression can be expected when using the ionic transition itself as a filter.
The results shown in this letter pave the way towards a whole range of new applications with the two most prominent ones being that
our device constitutes a monolithic source for quantum states of light at UV wavelengths, which may be further improved by specifically tailoring the SPDC process, 
and
that operating the device in the reverse direction, i.e. downconverting light from $369.5\,\mathrm{nm}$ to the telecommunications O-band, does not pose a fundamental problem. 
We therefore expect our device to be highly useful for quantum information tasks involving direct access to trapped ion systems in the ultraviolet spectral region.
\\~
\begin{acknowledgments}
\textit{Acknowledgements} We thank Stephan Krapick and Viktor Quiring for help in designing and producing the filtering etalon used in this work and Harald Herrmann for helpful discussions.
We acknowledge financial support provided by the Bundesministerium f\"ur Forschung und Bildung within the \textit{QuOReP} and \textit{Q.com-Q} framework.
\end{acknowledgments}

%

\end{document}